\documentclass[a4paper,aps,pra,showpacs,preprintnumbers,twocolumn]{revtex4}

\usepackage{graphics,graphicx,epsfig}
\usepackage[centertags]{amsmath}
\usepackage{amsfonts}
\usepackage{amssymb}
\usepackage[cp1251]{inputenc}
\usepackage[english]{babel}
\usepackage{url}
\inputencoding{cp1251}
\usepackage{graphicx}
\usepackage{amsthm,color}
\DeclareGraphicsExtensions{.pdf,.png,.jpg,.eps}
\usepackage{epstopdf}
\begin{document}

\def\BE{\begin{equation}}
\def\EE{\end{equation}}
\def\BEA{\begin{eqnarray}}
\def\EEA{\end{eqnarray}}
\def\BY{\begin{eqnarray}}
\def\EY{\end{eqnarray}}

\def\L{\label}
\def\nn{\nonumber}
\def\ds{\displaystyle}
\def\o{\overline}

\def\({\left (}
\def\){\right )}
\def\[{\left [}
\def\]{\right]}
\def\<{\langle}
\def\>{\rangle}

\def\h{\hat}
\def\hs{\hat{\sigma}}
\def\td{\tilde}

\def\k{\mathbf{k}}
\def\q{\mathbf{q}}
\def\r{\vec{r}}
\def\ro{\vec{\rho}}
\def\a{\hat{a}}
\def\b{\hat{b}}
\def\c{\hat{c}}
\def\h{\hat}

%-----------------------------------------------------------

 %\large

\title{ Quantum theory of a laser soliton}

%\vspace{1cm}
\author{Golubeva T. Yu.$^1$, Golubev Yu.M.$^1$, Fedorov S.V.$^2$,\; Nesterov L.A.$^{2\dag}$\footnotetext{\dag \; deceased}, Vashukevich E.A.$^1$, Rosanov N.N.$^{2,3,4}$}
\address{$^{1}$Saint-Petersburg State University,
$^{2}$Ioffe Institute, St. Petersburg,
 $^{3}$Vavilov State Optical Institute,  $^{4}$ITMO
University}
%\address{Saint-Petersburg State University, St Petersburg, 199034, Russia}
%\date{\today}

\begin{abstract}
The Heisenberg-Langevin equation for a spatial laser soliton in a wide-aperture laser with saturable absorption is constructed within the framework of consistent quantum electrodynamics. We discuss in detail the canonical variables for the generation field and the material two-level medium, consisting of centres providing amplification and absorption. It is assumed that laser generation evolves in time much more slowly than an atomic media. This assumption makes it possible to apply the adiabatic approximation and construct a closed equation for the amplitude of a laser field. Much attention is paid to the formulation of Langevin sources when deriving the equation since they play a decisive role in the formation of solitons' quantum statistical features. To provide an appropriate procedure for observing the quantum squeezing of a soliton, synchronization of laser generation by an external weak electromagnetic field is considered. Here we also present the results of the analysis of a classical laser soliton (neglecting the quantum fluctuations), which serves as the basis for further consideration of quantum effects.

\end{abstract}


\maketitle

\section{Introduction}
The dissipative optical solitons \cite{RosanovDOS} are of considerable interest since the realized dynamic balance of energy afflux and drain provides their increased stability compared to the more studied conservative solitons \cite{NovikovZakharov1984}. The extreme possibilities of information applications of dissipative solitons are dictated by quantum fluctuations of the field and the medium in which they propagate. The squeezed states with a noise level lower than that in the coherent state are theoretically predicted and experimentally realized for the radiation corresponding to conservative temporal solitons \cite{Carter1987, Rosenbluh1991, Spalter1998}. Similar studies of the quantum fluctuations of spatial conservative solitons were performed in \cite{Mecozzi1997, Lantz2004, Boyd1998}.

For spatial dissipative optical solitons, quantum effects have not been studied in such detail. The analysis of squeezed radiation of such solitons in a parametric oscillator scheme are summarized in the monograph \cite{OppoInKolobov2007}. In \cite{NesterovKiselevRosanov2009, NesterovVeretenovRosanov2015, NesterovVeretenov2015}, the Langevin quantum equation for spatial solitons in a wide-aperture passive nonlinear interferometer with coherent external radiation is derived. Obtained solution of this equation makes it possible to find the statistical dispersion of fluctuations of the average position and momentum of the soliton and the conditions for the squeezed states realization.

However, currently there are no studies of quantum fluctuations of spatial dissipative solitons in another scheme -- a wide-aperture laser with saturable absorption. Such a scheme, proposed in \cite{RosanovFedorov} (for an analysis of the subsequent studies of this scheme, see \cite{RosanovDOS}), compares favourably with the scheme of an interferometer with a high-intensity contrast in the centre of the soliton in comparison to its periphery. The development of the theory of quantum effects for such laser solitons is the main goal of this report. For generality and methodological purposes, we consider a more universal scheme of a wide-aperture laser with saturable absorption and holding weak radiation. This scheme is reduced to the simple laser scheme for zero intensity of the holding radiation, or to the scheme of a nonlinear interferometer with holding radiation when replacing the material equations for the medium.
%1. Розанов Н.Н. Диссипативные оптические солитоны. М.: Физматлит, 2011.

%2. Novikov S., Manakov S.V., Pitaevskii L.P., Zakharov V.E. Theory of Solitons. The Inverse Scattering Method., New York, Consultants Bureau, 1984.
%3. Carter S.J., Drummond P.D., Reid M.D., Shelby R.M. Phys. Rev. Lett. 1987. V. 58. P. 1841.
%4. Rosenbluh M. and Shelby R.M. Phys. Rev. Lett. 1991. V. 66. P. 153.
%5. Spalter S. et al. Opt. Express. 1998. V. 2. P. 77.
%6. Mecozzi A., Kumar P. J. Opt. B: Quantum Semiclass. Opt. 1998. V. 10. P. L21.
%7. Lantz E. et al. J. Opt. B. Quantum Semiclass. Opt. 2004. V. 16. P. S295.
%8. Nagasako E.M., Boyd R.W., Agarwal G.S. Opt. Express. 1998. V. 3. P. 171.
%9. Oppo G-L., Jeffers J. In: Quantum Imaging. Ed. M. Kolobov. Berlin: Springer, 2007. P. 13.
%10. Nesterov L.A., Kiselev Al.S., Kiselev An.S., Rosanov N.N. Opt. Spectr. V. 106 (4). P. 570-588, 2009.
%11. Nesterov L.A., Veretenov N.A., Rosanov N.N. Opt. Spectr. V. 118(5). P. 781–793, 2015.
%12. Nesterov L.A., Veretenov N.A., Rosanov N.N. Opt. Spectr. V. 118(5). P. 794–802, 2015.
%13. Rozanov N.N., Fedorov S.V. Optics and Spectroscopy. V. 72. Issue 6. P. 782-785, 1992.

\section{Hamiltonian for laser soliton}

\subsection{Physical model of a laser soliton}
It is assumed that there are two different media consisting of two-level atoms in an travelling wave optical cavity with volume $V=SL$ ($S$ is the cross-sectional area of the cavity, $L$ is the perimeter of the cavity). One of them is the active medium and it provides laser generation, and the other is the passive one, providing saturated absorption of the generation field. Fig. 1 shows the energy structure for the atoms of the active and passive media.
\begin{figure}[h]
\includegraphics[scale=0.4]{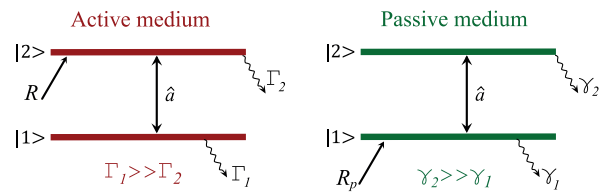} %{Fig1.eps}
\caption{ The schematic draw of the active and passive mediums.} \L{fig1}
\end{figure}

It can be seen that the upper level of the active medium is excited due to incoherent pump at a rate of $ R $. The passive medium is excited to the lower level at a rate of $ R_p $. Laser generation described by the Heisenberg operator $ \hat a $ and assumed to be resonant to both media. Each of the levels of these two media spontaneously relaxes to some third levels.

As for the optical cavity, it is a cavity of travelling quasiplane waves. One of the waves travelling along the $ z $ axis (along the cavity axis) is a laser wave. It is assumed that the transverse cavity size $ \sqrt S $ is much larger than its perimeter $ L $ so that the transverse modes form a continuous spectrum.

Initially (before generation), the atoms of both media considered to be evenly distributed inside the cavity. However, we will take into account the possibility of the appearance of transverse spatial inhomogeneity in the form of a laser soliton when constructing the theory.

\subsection{The amplitude of laser generation}

The intracavity generation field will be considered as quasi-plane quasimonochromatic wave. We can write down the electric field in the form
\BY
&&\hat E(\vec r,t)=i\(\frac{\hbar\omega_0
}{2\varepsilon_0L}\)^{1/2}\;e^{\ds ik_0z-i\omega_{0}\;t}\;\hat
a(z,\vec\rho,t),\nn\\
&& \vec r=(z,\vec\rho),\qquad \vec\rho=(x,y).\L{1}
\EY
The Heisenberg operator $\hat a(z, \vec\rho, t)$  is treated as a field amplitude operator, describing, in particular, a laser soliton. Since the optical frequency in this expression is distinguished as a separate factor, the amplitude $\hat a (z,\vec \rho, t)$ is slow. Here we will aim to obtain a closed equation for this quantity.

Since any field inside a cavity can be represented as
a linear superposition of plane eigenmodes of the cavity, the operator of the electromagnetic field takes the form:
\BY
&&\hat E(\vec r,t)=i\(\frac{\hbar}{2\varepsilon_0
V}\)^{1/2}\sum_{\vec k}\sqrt{\omega_{\vec k}}\;\hat a_{\vec
k}\;e^{\ds i\vec k\vec r-i\omega_{\vec k}t}, \nn\\
&& \[\hat a_{\vec k},\hat a^\dag_{\vec k^\prime}\]=\delta_{\vec
k\vec k^\prime}.\L{2}
\EY
Here by the index $ \vec k $ the set of indices $k_x, k_y, k_z $ is meant
\BY
&& k_x=\frac{2\pi}{\sqrt S}n_x,\qquad  k_y=\frac{2\pi}{\sqrt
S}n_y,\qquad k_z=\frac{2\pi}{L}n_z,\nn\\
&&n_x,n_y,n_z=0,\pm1,\pm2,\cdots,\L{3}
\EY
We represent the wave vector $ \vec k $ as the sum $ \vec k = \vec
k_z + \vec q $, where the quantity $ k_z $ numbers the longitudinal waves running along the axis $ z $, and the vector $ \vec q $ numbers the transverse waves.

Comparing the formulas (\ref{1}) and (\ref{2}) with each other, it is easy to obtain:
\BY
&&\hat a(z,\vec\rho,t)=\frac{1}{\sqrt{S}}\sum_{\vec k} \sqrt{{\omega_{\vec
  k}}/{\omega_0}}\;\hat a_{\vec k }\times\nn\\
&&e^{\ds i(k_z-k_0)z+ i\vec q\;\vec\rho-i(\omega_{\vec
k}-\omega_{0})t}.\L{4}
\EY

Further, we will assume that only one longitudinal mode $k_z=k_0$ engaged in laser generation. As can be seen, the laser amplitude ceases to depend on the coordinate $z$, and summation over the wave vector $\vec k$ turns into summation over the wave vector $\vec q$, when determining the amplitude. The decomposition (\ref{4}) will be further used to build the commutation relation $\[\hat a(\vec\rho, t), \hat a^\dag (\vec\rho\;^\prime, t) \] $.

\subsection{Collective variables for active and passive medium}

As already noted, the cavity is evenly filled with a medium consisting of two-level atoms. These atoms form two independent ensembles. One of them provides laser generation, and the other provides saturated absorption of laser generation. We suppose that all atoms are motionless and do not interact with each other. For simplicity, we assume that the transition frequencies are the same for different atoms. Each two-level atom of the active medium is described by three Heisenberg amplitudes: the coherence $\hat \sigma^j $,
the population of the upper level $\hat\sigma_2^j$ and the population of the lower level $\hat \sigma_1^j $. Here the index $j$ numbers the atoms of the active medium. Similarly, for a passive medium, the variables $ \hat \pi^j $, $\hat \pi^j_2$ and $ \hat \pi_1^j$  introduced. Individual amplitudes are written as projection operators
\BY
&&\hat\sigma^j=|1>_{jj}<2|,\quad\hat\sigma_{1}^j=|1>_{jj}<1|,\quad \hat\sigma_{2}^j=|2>_{jj}<2|,\nn\\
&&\hat\pi^j=|1>_{jj}<2|,\quad\hat\pi_{1}^j=|1>_{jj}<1|,\quad
\hat\pi_{2}^j=|2>_{jj}<2|.\nn\\\L{5}
\EY
and there are commutation relations
\BY
&&\[\hat\sigma^j(\vec\rho_j,t),\hat\sigma^{j\dag}(\vec\rho_j,t)\]=\hat\sigma^j_{1}(\vec
\rho_j,t)-\hat\sigma^j_{2}(\vec \rho_j,t)\L{6}\\
&&\[\hat\sigma^j_1(\vec\rho_j,t),\hat\sigma^j(\vec\rho_j\;^\prime,t)\]=
\hat\sigma^j(\vec\rho_j,t),\L{7}\\
&&\[\hat\sigma^j_2(\vec\rho_j,t),\hat\sigma^j(\vec\rho_j,t)\]=
-\hat\sigma^j(\vec\rho_j,t).\L{8}
\EY
\BY
&&\[\hat\pi^j(\vec\rho_j,t),\hat\pi^{j\dag}(\vec\rho_j,t)\]=\hat\pi^j_{1}(\vec
\rho_j,t)-\hat\pi^j_{2}(\vec \rho_j,t)\L{9}\\
&&\[\hat\pi^j_1(\vec\rho_j,t),\hat\pi^j(\vec\rho_j\;^\prime,t)\]=
\hat\pi^j(\vec\rho_j,t),\L{10}\\
&&\[\hat\pi^j_2(\vec\rho_j,t),\hat\pi^j(\vec\rho_j,t)\]=
-\hat\pi^j(\vec\rho_j,t).\L{11}
\EY
Here we suggested that the medium is homogeneous in the longitudinal
direction and does not depend on the spatial variable $z$, just as it was for the field.

Further we will not follow individual variables, but collective ones, which are defined as linear superpositions of individual variables in the form
 \BY
&&\hat\sigma(\vec \rho,t)=
-i\sum_j\hat\sigma^j(\vec\rho_j,t)\;\Theta(t-t_j)\delta^2(\vec
\rho-\vec \rho_j)\times\nn\\
&&e^{\ds -ik_0z_j+i\delta_a t},\L{12}\\
&&\hat\sigma_{1}(\vec \rho,t)=
\sum_j\hat\sigma_{1}^j(\vec\rho_j,t)\; \Theta(t-t_j)\delta^2(\vec
\rho-\vec
\rho_j),\L{13}\\
&&\hat\sigma_{2}(\vec \rho,t)=
\sum_j\hat\sigma_{2}^j(\vec\rho_j,t)\; \Theta(t-t_j)\delta^2(\vec
\rho-\vec \rho_j).\L{14}
\EY
Here $ \delta_a = \omega_0- \omega_a $ is the shift of the laser frequency regarding the atomic frequency at the laser transition.

Collective variables of passive media are defined similarly:
\BY
&&\hat\pi(\vec \rho,t)=
-i\sum_j\hat\pi^{j}(\vec\rho_{j},t)\;\Theta(t-t_j)\delta^2(\vec
\rho-\vec \rho_j)\times\nn\\
&&e^{\ds -ik_0z_j+i\delta_p },\L{15}\\\nn\\
&&\hat\pi_{1}(\vec \rho,t)= \sum_j\hat\pi_{1}^j(\vec\rho_j,t)\;
\Theta(t-t_j)\delta^2(\vec \rho-\vec
\rho_j),\L{16}\\
&&\hat\pi_{2}(\vec \rho,t)= \sum_j\hat\pi_{2}^j(\vec\rho_j,t)\;
\Theta(t-t_j)\delta^2(\vec \rho-\vec \rho_j).\L{17}
\EY
Here $ \delta_p = \omega_0- \omega_p $ is the shift of the laser frequency regarding the atomic frequency of the passive medium.

The Hamiltonians of the interaction of the laser field $ \hat a (\vec \rho, t) $ with both media under the dipole approximation and the approximation of slowly rotating amplitudes have the form
\BY
 \hat V_a=\!\!\!\!\!\!&&\hbar g\int_S\;d^2\rho\[\hat\sigma(\vec \rho,t)\;\hat
a^\dag(\vec\rho,t)+\hat\sigma^\dag(\vec \rho,t)\;\hat
a(\vec\rho,t)\],\;\L{18}\\
 \hat V_p=\!\!\!\!\!\!&&\hbar g_p\int_S\;d^2\rho\[ \hat\pi(\vec \rho,t)\hat
a^\dag(\vec\rho,t)+\hat\pi^\dag(\vec \rho,t)\;\hat a(\vec\rho,t)\].\;\;\L{19}
\EY
Here the interaction constants are expressed explicitly in terms of the off-diagonal matrix elements of the dipole moments $ \mu $ (for the active medium) and $ \mu_p $ (for the passive medium) in the form
\BY
&& g=\mu\(\frac{\omega_0}{2\varepsilon_0 \hbar L}\)^{1/2},\quad
g_p=\mu_p\(\frac{\omega_0}{2\varepsilon_0 \hbar L}\)^{1/2}. \L{20}
\EY

\section{The initial system of equations for the field and the matter}

\subsection{Commutation relations for collective atomic variables and field amplitude}

To write the equations for operator quantities, it is necessary to know the corresponding commutation relations. Firstly, let us discuss the field variable. We rewrite the equality (\ref {4}) taking into account that only one laser mode among the longitudinal waves is included in generation. To do this, we replace the summation over $ \vec k $ with the summation over $ \vec q $. In this case, we obtain
 \BY
&&\hat a( \vec\rho,t)=\frac{1}{\sqrt{S}}\sum_{\vec q}
\sqrt{{\omega_{\vec
  q}/\omega_0}}\;\hat a_{\vec q }\times\nn\\
  &&e^{\ds   i\vec
q\;\vec\rho-i(\omega_{\vec q}-\omega_{0})t}.\L{21}
\EY
Consequently, equality for the commutation relation can be written in the following form:
\BY
&&\[\hat a(\vec\rho,t),\hat a^\dag(\vec\rho\;^\prime,t)\]=\frac{1}{S}\sum_{\vec
q}{\omega_{\vec q}}/{\omega_0}\: e^{\ds i\vec q(\vec
\rho-\vec\rho\:^\prime)}.\L{22}
\EY
We take into account that the transverse cavity size $ \sqrt S $ is so large that the transverse modes form a continuous spectrum. This allows us to replace the summation over the transverse modes with integration according to the rule
\BY
&&\sum_{\vec q}\cdots\to\frac{S}{(2\pi)^2}\int d^2q\cdots\L{23}
\EY
We can represent an arbitrary wave vector as the sum of the wave vectors of the longitudinal and transverse waves $ \vec k = \vec k_0 + \vec q $. Consequently, the $ k^2 = k_0^2 + q^2 $, and then under the paraxial approximation $ k_0 \gg q $  the following formula can be written
\BY
&&\frac{\omega_{\vec q}}{\omega_0}\approx
1+\frac{q^2}{2k_0^2}.\L{24}
%9
\EY
Replacing the summation in the formula (\ref{22}) with integration according to the rule (\ref{23}), we obtain after integration the expression
\BY
&&\[\hat a(\vec\rho,t),\hat
a^\dag(\vec\rho\;^\prime,t)\]=\tilde\delta(\vec \rho-\vec
\rho\;^\prime).\L{25}
\EY
Here we denote
\BY
&&\tilde\delta(\vec \rho-\vec\rho
\;^\prime)=\(1-\frac{1}{2k_0^2}\Delta_\perp\)\delta^2(\vec\rho-\vec\rho\;^\prime),\nn\\
&&\Delta_\perp=\frac{\partial^2}{\partial
x^2}+\frac{\partial^2}{\partial y^2}.\L{26}
\EY
Now let us discuss the commutation relations for atomic variables. To do this we use the definition of collective variables (\ref{12}) - (\ref{14}) and the commutation relations for individual variables (\ref{6}) - (\ref{8}). Then we can write the following equalities for the active medium
\BY
&&\[\hat\sigma(\vec\rho,t),\hat\sigma^\dag(\vec\rho\;^\prime,t)\]=\[\hat\sigma_{1}(\vec\rho,t)-\hat\sigma_{2}(\vec\rho,t)\]
\times\nn\\
&&\delta^2(\vec\rho-\vec\rho\;^\prime),\L{27}\\
&&\[\hat\sigma_1(\vec\rho,t),\hat\sigma(\vec\rho\;^\prime,t)\]=
\hat\sigma(\vec\rho,t)\;\delta^2(\vec\rho-\vec\rho\;^\prime),\L{28}\\
&&\[\hat\sigma_2(\vec\rho,t),\hat\sigma(\vec\rho\;^\prime,t)\]=
-\hat\sigma(\vec\rho,t)\;\delta^2(\vec\rho-\vec\rho\;^\prime).\L{29}
\EY

The definition of the collective variable $ \hat\pi $ exactly coincide with that of $\hat\sigma $, so the commutation relations for  $ \hat\pi $ will also coincide with relations for $\hat\sigma $
\BY
&&\[\hat\pi(\vec\rho,t),\hat\pi^\dag(\vec\rho\;^\prime,t)\]=\[\hat\pi_{1}(\vec\rho,t)-\hat\pi_{2}(\vec\rho,t)\]\times\nn\\
&&\delta^2(\vec\rho-\vec\rho\;^\prime),\L{30}\\
&&\[\hat\pi_1(\vec\rho,t),\hat\pi(\vec\rho\;^\prime,t)\]=
\hat\pi(\vec\rho,t)\;\delta^2(\vec\rho-\vec\rho\;^\prime),\L{31}\\
&&\[\hat\pi_2(\vec\rho,t),\hat\pi(\vec\rho\;^\prime,t)\]=
-\hat\pi(\vec\rho,t)\;\delta^2(\vec\rho-\vec\rho\;^\prime).\L{32}
\EY

\subsection{Heisenberg-Langevin system of equations}

In the presence of the Hamiltonian and commutation relations, we allowed to use standard recipes of quantum theory and write Heisenberg equations for operators of field and atomic variables. As is clear, these equations cannot be considered completely satisfactory, since such important circumstances as field and substance relaxation are not taken into account in this equations. The factor of excitation of the medium is also do not consedered.

To fix the situation we should proceed as follows. The actual terms of relaxation (leaving the system) and excitation (coming from outside) are phenomenologically introduced into the equations. This action is completely legitimate in calculations in classical electrodynamics. But in the framework of quantum theory it should be supplemented, according to the fluctuation-dissipation theorem, by the introduction of Langevin forces. The resulting new equations are usually called the Heisenberg-Langevin equations. It should be noted that we called this approach phenomenological, but, in fact, this is not so, as this approach could be easily justified within the framework of quite adequate physical models \cite{Scully}.

The obtained Heisenberg-Langevin equations are quite suitable for physical analysis. However, to simplify the mathematical expressions, let us replace the operator quantities by c-numbers . To do this, it is enough to assume that we will be interested only in such measurement procedures that are associated with the observation of normally ordered means. According to, for example, \cite{Scully} normal ordering of operators is defined as $ \hat a^\dag \hat\sigma^\dag\hat \sigma_2\hat\sigma_1\hat\sigma\hat a $ or $ \hat a^\dag \hat \pi^\dag\hat\pi_2\hat\pi_1\hat\pi\hat a $.

The transition to c-numbers could be done through the following steps. First of all, it is necessary to put the product of operator quantities in the normal order using commutation relations in every term of the Heisenberg-Langevin equations. Then all operators (including Langevin operators) must be turned into c-numbers. To do this, simply remove the "hats" \;, indicating the operator nature of the quantities. Moreover, the properties of Langevin sources change significantly. The work \cite{Davidovich1996} investigated in detail how to find new correlation functions for these sources.

The equations in the c-number representation can be written as
\BY
&&\frac{\partial a}{\partial t}-\frac{ic}{2k_0}\;\Delta_\perp
 a=-\kappa/2\; a+g \sigma+g_p \pi,\L{33}\\
&&\frac{\partial\sigma}{\partial
t}=-(\Gamma-i\delta_a)\sigma+g(\sigma_2-\sigma_1) a+ F,\L{34}\\
&&\frac{\partial\sigma_1}{\partial
t}=-\Gamma_1\sigma_1+g(\sigma^\ast\hat
a+\sigma a^\ast )+ F_1,\L{35}\\
&&\frac{\partial\sigma_2}{\partial
t}=R_a-\Gamma_2\sigma_2-g(\sigma^\ast
a+ a^\ast\sigma)+ F_2,\L{36}\\
&&\frac{\partial\pi}{\partial
t}=-(\gamma-i\delta_p)\pi+g_p(\pi_2-\pi_1)
 a+ G,\L{37}\\
&&\frac{\partial\pi_1}{\partial t}=R_p-\gamma_1\pi_1+g_p(\pi^\ast
a+
 a^\ast\pi)+ G_1,\L{38}\\
&&\frac{\partial\pi_2}{\partial t}=-\gamma_2\pi_2-g_p(\pi^\ast
a+\pi a^\ast)+ G_2.\L{39}
\EY
These equations are very similar to operator ones, but "operator hats"\;are removed everywhere.

As can be seen from the equations, we took into account the relaxation of the active medium from the upper laser level at a speed of $ \Gamma_2 $ and the excitation of this level at a speed of $ R_a $ (pump). The lower laser level decays at a speed of $\Gamma_1 $. Spontaneous decay from the upper laser level to the lower one is not considered.

In the same way, the spontaneous relaxation of the passive medium from the upper to the lower level is not considered. A passive medium is excited through the lower level at a speed of $ R_p $ and relaxation from the lower level occurs at a speed of $ \gamma_1 $. Decay rate from the top-level denoted as $ \gamma_2 $.

Langevin sources for the active medium in c-number
representation are defined by the following equalities.  First, we write the general equality
\BY
&&\langle F_i^\ast(\vec\rho,t)
F_j(\vec\rho\;^\prime,t\;^\prime)\rangle=\langle F_i^\ast
F_j\rangle\;
\delta(t-t^\prime)\delta^2(\vec\rho-\vec\rho\;^\prime),\L{}\nn
\EY
In this case, nonzero correlators are written as
\BY
&&\langle F^\ast F\rangle=\Gamma_1\o\sigma_2+R_a,\L{40}\\
&&\langle F F\rangle=2g\;\o{\sigma a},\L{41}\\
&&\langle F_1
F\rangle=\Gamma_1\o\sigma,\L{42}\\
&&\langle F_2 F_2\rangle=\Gamma_2\o\sigma_2 +R_a(1-s_a)- g\(\o{
a^\ast\sigma}+\o{a\sigma^\ast}\),\L{43}\\
&&\langle F_1 F_1\rangle=\Gamma_1\o\sigma_1-
g\(\o{ a^\ast\sigma}+\o{a\sigma^\ast}\),\L{44}\\
&&\langle F_2 F_1\rangle=
g\(\o{a^\ast\sigma}+\o{a\sigma^\ast}\).\L{45}
\EY
We took these correlation relations from \cite{Davidovich1996}, generalizing it to the physical system with transverse inhomogeneity.

Here it is necessary to pay attention to the parameter $ 0 \leq s_a \leq1 $, which enables us to consider not only the average excitation rate of the upper laser level $ R_a $, but also to follow the fluctuation of the excitations. In the extreme case $ s_a = 0 $, atoms are excited independently of each other and randomly in time (Poissonian pump). And for $ s_a = 1 $, there is strictly regular pump without noise (sub-Poissonian pump).

We can write down the correlation properties of sources for a passive medium
\BY
&&\langle G_i^\ast(\vec\rho,t)
G_j(\vec\rho\;^\prime,t^\prime)\rangle=\langle G_i^\ast
G_j\rangle\;\delta(t-t^\prime)
\delta^2(\vec\rho-\vec\rho\;^\prime),\L{}\nn\\\nn\\
&&\langle G^\ast
G\rangle=R_p+\gamma_1\o\pi_2-\gamma_1\o\pi_1+\gamma_2\o\pi_2+\nn\\
&&+2g_p\(\o{
a^\ast\pi}+\o{a\pi^\ast}\),\L{46}\\
&&\langle G G\rangle=
2g_p\;\o{\pi a},\L{47}\\
&&\langle G_2
G\rangle=\gamma_2\o\pi,\L{48}\\
&&\langle G_1 G_1\rangle =\gamma_1\o\pi_1+R_p(1-s_p)-\nn\\
&&-g_p\(\o{ a^\ast\pi}+\o{a\pi^\ast}\),\L{49}\\
&&\langle G_2 G_2\rangle=\gamma_2\o\pi_2-
g_p\(\o{ a^\ast\pi}+\o{a\pi^\ast}\),\L{50}\\
&&\langle G_1 G_2\rangle=
g_p\(\o{a^\ast\pi}+\o{a\pi^\ast}\).\L{51}
\EY
The meaning of the $ s_p $ parameter is exactly the same as $ s_a $, but with respect to the lower level of the passive medium. The correlation properties of Langevin sources for a passive medium was calculated following the recipes \cite{Davidovich1996}.

In \cite{GolubevFedRos2019}, the correlation relations for the passive medium (\ref{46}) - (\ref{51}) were calculated incorrectly.
\section{A quantum equation for the field amplitude under the adiabatic approximation}

The system of seven equations (\ref{33}) - (\ref{39}) completely describes the laser system, including the soliton case. It, in principle, make it possible to obtain all the necessary information about the generation field formally represented by the amplitude $ a (\vec \rho, t) $. However, we will make further efforts here to build a closed equation for this amplitude. It can be done under the assumption that the field variable evolves much more slowly than atomic ones. This situation could be described with the following inequalities
\BY
&&\kappa\ll\Gamma,\Gamma_1,\Gamma_2,\qquad \kappa\ll\gamma,\gamma_1,\gamma_2.\L{52}
\EY
As is known, in this case, it is possible to exclude atomic variables adiabatically from consideration and construct a closed equation for a field variable.

To avoid overly complex formulas, we require the fulfillment of the following inequalities
\BY
&&\Gamma_2\ll\Gamma_1, \qquad \gamma_1\ll\gamma_2.\L{53}
\EY
These requirements provide inversion of the population of the upper level of the active medium and the lower level of the passive environment. For simplicity, we also assume that $ \delta_a = 0, \delta_p = 0 $, that is, the frequency of the laser field coincides with the mode frequency and with the frequencies of atomic transitions.

Assuming  that in equations (\ref {33}) - (\ref {39}) all time derivatives to be equal to zero, we obtain two independent systems of algebraic equations. One of them for the active medium in the form
\BY
&&\Gamma\sigma=g(\sigma_2-\sigma_1) a+ F,\L{54}\\
&&\Gamma_1\sigma_1=g(\sigma^\ast\hat
a+\sigma a^\ast )+ F_1,\L{55}\\
&&\Gamma_2\sigma_2=R_a-g(\sigma^\ast a+ a^\ast\sigma)+ F_2,\L{56}
\EY
and the other is for passive medium
\BY
&&\gamma\pi=g_p(\pi_2-\pi_1)
 a+ G,\L{57}\\
&&\gamma_1\pi_1=R_p+g_p(\pi^\ast a+
 a^\ast\pi)+ G_1,\L{58}\\
&&\gamma_2\pi_2=-g_p(\pi^\ast a+\pi a^\ast)+ G_2.\L{59}
\EY
We solve these equations for the atomic variables, assuming that the laser field amplitude remains unchanged in the adiabatic approximation. As a result, the variables $ \sigma $ and $ \pi $ will be expressed through the laser amplitude $ a (\vec \rho, t) $. We substitute these solutions into the equation (\ref{33}) and thereby obtain the closed equation for this laser amplitude in the form
\BY
&&\frac{\partial a}{\partial t}-\frac{ic}{2k_0}\;\Delta_\perp
 a=-\frac{1}{2}\kappa (a-a_{in}e^{-i\nu_{in}t}) +\nn\\
 &&+\frac{1}{2}\(\frac{A}{1+\beta|a|^2}-\frac{A_p}{1+\beta_p|a|^2}\)a+\Phi+\Phi_p.\L{60}
\EY
In the equation $A=\beta R_a$ and $A_p=\beta_p R_p$ are the linear gain of the active medium and the absorption of the passive medium. The values $\beta={4g^2}/{(\Gamma_1 \Gamma_2)}$ and $\beta_p={4g_p^2}/{(\gamma_1\gamma_2)} $ determine the nonlinear properties of the active medium and, accordingly,the passive one. The inverse quantities $\beta^{-1}, \; \beta^{-1}_p $ have the meaning of the number of photons saturating the corresponding atomic transition.

When writing the equation (\ref{60}), we took into account the fact that in order to observe quantum features, it is necessary to suppress phase diffusion in the laser. Laser radiation can be synchronized by a weak external field in a coherent state. We denote the amplitude of the holding field in the form $ a_{in} $ and assumed that the frequency this field is shifted from the current mode frequency by $ \nu_{in} $.

\section{The Langevin sources in adiabatic c-number theory}

In order to calculate the Langevin sources $ \Phi $ and $ \Phi_p $ in equation (\ref {60}), we must use the correlation relations (\ref{40}) - (\ref{51}). However, the latter turn out to be undetermined, since they depend on unknown quantities $ \o \sigma $, $ \o \sigma_1 $, $ \o \sigma_2 $ and $ \o \pi $, $ \o \pi_1 $, $ \o \pi_2 $. To find these quantities, we use the equalities (\ref{54}) - (\ref{56}) and (\ref{57}) - (\ref{59}). After averaging them, two closed algebraic systems of equations arise with respect to the indicated quantities. Solving the equations, we obtain for the active medium
\BY
&&\o\sigma_1=\frac{R_a}{\Gamma_1}\;\frac{I}{1+I},\qquad\o\sigma_2=\frac{R_a}{\Gamma_2}\;\frac{1}{1+I},\nn\\
&&g\o\sigma=\frac{1}{2} \frac{\beta R_aa}{1+I},\qquad I=\beta|a|^2,\L{61}
\EY
and for the passive one
\BY
&&\o\pi_1=\frac{R_p}{\gamma_1}\;\frac{1}{1+I_p},\qquad\o\pi_2=\frac{R_p}{\gamma_2}\;\frac{I_p}{1+I_p},\nn\\
&&g_p\o\pi=-\frac{1}{2} \frac{\beta_pR_pa}{1+I_p},\qquad I_p=\beta_p|a|^2.\L{62}
\EY
Substituting this into the formulas (\ref{40}) - (\ref{51}), we obtain fully defined correlation properties of the original Langevin sources. Now we have the opportunity to explicitly write the sources $ \Phi $ and $ \Phi_p $.

In the adiabatic equation, the Langevin force is represented by two terms. The first of them is formed by the active medium and turns out to be a combination of the initial sources $ F, F_1, F_2 $, and the second is the combination of the sources $ G, G_1, G_2 $, that is, it is formed by the passive medium. It turns out to be convenient for calculations to write down sources in the form
\BY
&&\Phi= \frac{1}{2}\frac{\beta
a}{1+\beta|a|^2}\;\xi_{2}+\xi_a,\L{63}\\
&&\Phi_p= -\frac{1}{2}\frac{\beta_p
a}{1+\beta_p|a|^2}\;\xi_{p1}+\xi_{pa},\L{64}
\EY
Here the parameters $\xi_a, \xi_2$ determine the contributions from the initial sources $ F, F_1, F_2 $ in the form
\BY
&&\xi_{a}= -\frac{2g^2a}{\Gamma_1^2}F_1+\frac{2g}{\Gamma_1}\;F ,\L{65}\\
&&\xi_{2}=-\frac{2g}{\Gamma_1}(aF^\ast+a^\ast F)+
\frac{4g^2}{\Gamma_1^2}|a|^2F_1+ F_2.\L{66}
\EY
The parameters $\xi_{pa}, \xi_{p1} $ determine the contributions from the sources $ G, G_1, G_2 $
\BY
&&\xi_{pa}= \frac{2g_p^2a}{\gamma_2^2}G_2+\frac{2g_p}{\gamma_2}\;G
,\L{67}\\
&&\xi_{p1}=\frac{2g_p}{\gamma_2}(aG^\ast+a^\ast G)+
\frac{4g_p^2}{\gamma_2^2}|a|^2G_2+ G_1.\L{68}
\EY
Now we are able to calculate the mean values we need. It is easy to obtain the following nonzero moments
\BY
&&\langle
\xi_i^\ast(t,\vec\rho)\xi_j(t,\vec\rho\;^\prime)\rangle=\langle
\xi_i^\ast\xi_j\rangle\delta(t-t^\prime)\delta^2(\vec\rho-\vec\rho\;^\prime),\L{}\nn\\\nn\\
&&\langle\xi_2\xi_2\rangle=2R_a(1-s_a/2),\qquad\langle
\xi_{p1}\xi_{p1}\rangle=2R_p(1-s_p/2 ),\nn\\
&&\langle \xi^\ast_a\xi_a\rangle=\frac{\beta
R_a}{1+\beta|a|^2},\qquad\langle\xi_a\xi_2\rangle=-\;\frac{\beta
R_aa}{1+\beta|a|^2}.\L{69}
\EY

There is an essential circumstance lies in the fact that the physical parameters $ (\gamma_1 / \gamma_2 I_p) $, $ (\gamma_1 / \gamma_2) $, $ (\Gamma_2 / \Gamma_1) $ were chosen by us as extremely small, when calculating Langevin sources. At the same time, for a passive medium, it still remains possible to consider the saturation conditions that occurs if $I_p\gg1$.

Now we can write the correlation relations for the sources $ \Phi $ and $ \Phi_p $

\BY
&&  \langle \Phi^2\rangle=-\frac{1 }{2}\(\frac{\beta
a}{1+I}\)^2R_a(1+s_a/2)\L{70},\\
&& \langle|\Phi|^2\rangle=\frac{\beta
R_a}{1+I}-\frac{1}{2}\;\(\frac{\beta
|a|}{1+I}\)^2R_a(1+s_a/2),\L{71}\\
&& \langle \Phi_p^2\rangle=\frac{1 }{2}\;\(\frac{\beta_p a
}{1+I_p}\)^2R_p(1-s_p/2),\L{72}\\
&&\langle|\Phi_p|^2\rangle=\frac{1
}{2}\;\(\frac{\beta_p|a|}{1+I_p}\)^2R_p(1-s_p/2).\L{73}
\EY
Thus, the theory of laser soliton is completely formulated in the form of the Heisenberg-Langevin equation (\ref{60}).
\section{Classical theory of a driven laser with saturable
absorber}
{\it Governing equations}. Further, we intend to investigate the quantum features of the soliton close to the classical one. Therefore, in this section, we describe the main properties of the classical soliton.

Neglecting quantum fluctuations, the equation (\ref{60}) is written in the following dimensionless form, which coincides with that adopted in our previous works at $ {\cal E}_{in} = 0 $, see \cite{RosanovDOS,RosanovFedorov}:
\BY
&&\frac{\partial E}{\partial t} - i\Delta_{\perp} E = E_{in} +i\theta E +f\left( \left| E \right|^2 \right) \;E. \L{CL1}
\EY
Here, the time $t$ is normalized to linear amplitude losses, the transverse coordinates $\vec\rho =( x, y)$ -- to the width of the Fresnel zone, the intensity $I = |E|^2$ -- to the saturation intensity of the absorber, $ b = {\beta_p} / \beta $ is the ratio of the saturation intensities of the active and passive centers, $ \nabla_\bot^2 = \nabla_{\vec\rho}^2 $ (dashes in (\ref{CL1}) are omitted):
\BY
t = t'/\left( {\kappa /2} \right),\quad \vec\rho  = \vec\rho\;^\prime/\sqrt {c/\left( {\kappa {k_0}} \right)} , \\
a\left( {t,\vec\rho } \right) = E\left( {t\;^\prime,\vec\rho\;^\prime} \right){e^{ - i{\nu _{in}}t}}/\sqrt {{\beta _p}}.	\L{CL2}
\EY
For dimensionless gain and absorption coefficients in the center of the lines $ {g_0} $ and $ {a_0} $ and the total nonlinear gain $ f \left(I\right) $ we use the notation \cite{RosanovDOS, RosanovFedorov}:
\BY
{g_0} = A/\kappa,\quad  {a_0} = {A_p}/\kappa ,\\
 f\left( I \right) =  - 1 - \frac{{{a_0}}}{{1 + I}} + \frac{{{g_0}}}{{1 + I/b}}.	\L{CL3}
\EY
Here it was more convenient for us to consider the frequency of the external signal (holding radiation) as the carrier frequency and introduce the frequency mismatch between it and the frequency of the longitudinal cavity mode $\theta=\left({{\nu_{in}} - {\nu_0}} \right) / \left({\kappa / 2} \right) $. The mode frequency corresponds to the centers of the gain and absorption lines. Generally speaking, for $ \theta \neq 0 $ the function $ f (I) $ becomes complex. But if the frequency shift is much smaller than the line width, then the imaginary part of $ f $ is small and can be neglected. In numerical calculations, fixed values of the parameters $b = 10 $, $ {a_0} = 2 $, $ {g_0} = 2.08 $ for transversely one-dimensional and $ {g_0} = 2.11 $ for two-dimensional geometry are used.

{\it Stationary homogeneous lasing}. First, we would consider spatially homogeneous stationary solutions of (\ref{CL1}), assuming, without loss of generality, $ {E_{in}}> 0 $. For this solutions, setting $ E = \sqrt I {e^{i \varphi}} $, we reduce the equation (\ref{CL1}) to
\BY
{e^{ - i\varphi }}{E_{in}}/\sqrt I  =  - i\theta  - f\left( I \right).	\L{CL4}
\EY
Therefore there are two solutions with exact resonance ($\theta=0 $): inphase, $\varphi = 0$, and antiphase, $\varphi=\pi$. Fig. \ref{fig2} (a) shows the hysteresis dependence of the intensity of the homogeneous regime $ I $ on the intensity of the holding radiation $ {I_{in}} = |{E_{in}}{|^2} $ at a constant value of $ {g_0} $ for the two values of the detunings $ \theta $. The stability of solutions (\ref{CL4}) is checked by calculating the growth increment of small perturbations in the framework of linearized equations (\ref{CL1}). For the selected parameters, the lower and upper branches of the hysteresis curves in in-phase mode correspond to stable regimes, and all antiphase modes are unstable.
\begin{figure}[h]
\includegraphics[scale=0.65]{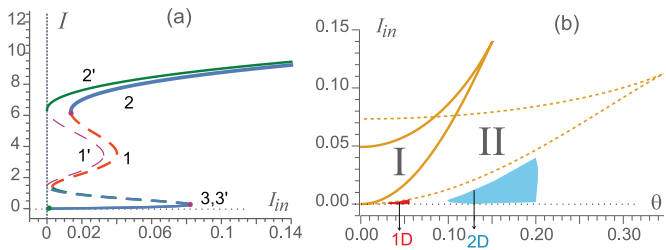}
\caption{ %Рис. 2.
 (a): Hysteretic dependence of the intensity of homogeneous laser generation on the intensity of the holding radiation at a gain of ${g_0} = 2.08 $. Curves 1, $ 1^\prime $ correspond to antiphase mode, and curves 2 and 3 correspond to in-phase mode. For curves 1, 2, 3: $\theta = 0.047 $, for $ 1^\prime $, $ 2^\prime $, $ 3^\prime $: $ \theta = 0 $. Dashed and dotted lines indicate unstable solitons, and continuous - stable ones. (b): Regions of bistability of in-phase (I) and antiphase (II) homogeneous generation modes and frequency locking regions of the 1D (red) and 2D (blue)  soliton ${g_0} = 2.08 $ (1D), $ 2.11 $ (2D) . The lower boundary of the locking regions corresponds to the incident beam intensity $ {I_ {in}}\approx 4\times {10^{-6}}$.
 \L{fig2}
 }
\end{figure}

{\it Synchronized solitons}. There are various types of spatial solitons corresponding to transversely localized generation with a certain frequency $ {\nu _s} $ in a laser with a saturable absorber without holding radiation \cite{RosanovDOS, RosanovFedorov}. The frequency is the eigenvalue of the problem
\BY
\Delta_{\perp} A + \left( \nu_s + \theta \right) = i f\left( \left| A \right|^2 \right) A,	\L{CL5}
\EY
arising from the substitution of $ E = A (\vec {\rho}) \exp {(- i \nu_s t)} $ into the master equation (\ref{CL1}). With very weak holding radiation, the field inside the cavity can be considered as a superposition of the soliton field and the field corresponding to stable homogeneous generation on the holding radiation frequency. If this frequency is not close to the soliton frequency, the total field is weakly modulated in time with the difference frequency $ {\nu_s} - {\nu_ {in}} $. Therefore, for the synchronization of a soliton by holding radiation, it is necessary to ensure that the intensity of the holding radiation exceeds a certain critical value, which depends on the frequency detuning $\theta $.
Fig. \ref{fig2} (b) shows the parameters' domains in which such synchronization is achieved in the transverse one-dimensional and two-dimensional cavity geometry. It can be seen from the figure that these domains are out of the bistability regions of homogeneous generation regimes. Calculations show that not only synchronization but also the localization of the structure disappears when the intensity of the holding radiation exceeds some other critical value. In this case, there is a consistent expansion of the region of either low-intensity homogeneous generation over the entire laser aperture or generation with beats, in dependence on the frequency detuning. It should be noted that with the optimal choice of the frequency detuning, holding radiation with a very low intensity $ {I_{in}} \approx 4 \times{10^{-6}} $ is sufficient to achieve synchronization.
The transverse profile of the synchronized soliton is shown in Fig. \ref{fig3} also for two geometry variants. At the periphery of the soliton, the field tends to the field of homogeneous generation considered above. The corresponding background is very small ($ | E | \approx 0.02 $). At the same time, spatial beats are observed in the peripheral part, which can be interpreted as the result of interference between the fields of the soliton itself and the field of homogeneous generation. In the central part of the soliton, significant changes in the profile of the soliton, in comparison with the variant of the absence of holding radiation, do not occur.
\begin{figure}[h]
\includegraphics[scale=0.65]{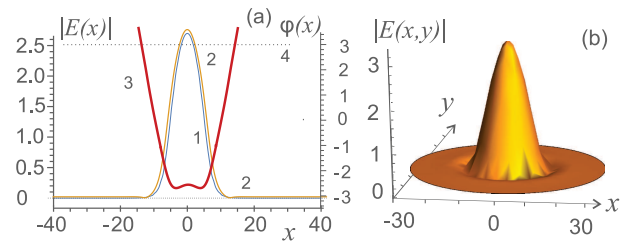}
\caption{ %Рис. 3.
Amplitude profiles for the one-dimensional (a) and two-dimensional (b) synchronized solitons. Curve 1 is a steady-state laser soliton without a holding beam, 2 is a synchronized soliton for $ {E_{in}} = 0.02 $, $ \theta = 0.043$, 3 is the phase profile of the synchronized soliton, 4 is the amplitude of the homogeneous generation regime. $ {g_0} = 2.08 $ (a), $ 2.11 $ (b).
} \L{fig3}
\end{figure}

\section{Conclusion}

The presented report is one of a series of works aimed on studying the dynamic and quantum-statistical properties of the laser soliton. Our first publication \cite{GolubevFedRos2019} could be considered as preliminary and was devoted to laser generation with the presence of saturated absorption. In that case, no spatial structure was taken into account, that is,  our consideration was not directly related to the laser soliton problem. Nevertheless, this allowed us to evaluate important methodological features of laser generation in the presence of an absorbing medium for their further use for soliton analysis.

On this methodological basis, in the present work, the non-linear Heisenberg-Langevin equation in partial derivatives was obtained. This equation describes the formation of the soliton and its statistics. It was assumed that the system under consideration can be described in the framework of the adiabatic approximation. This approximation could be achieved if the evolution of the laser field is slow in comparison with that of both (active and passive) atomic media.

Thus, at this stage, we make it possible to comprehensively describe the system in the future. All questions related to the quantum-statistical description can be explained when solving the obtained basic equation. In this work, we performed a numerical analysis of the solution of the equation in the classical limit. The main results of this analysis are as follows. For a wide-aperture driven laser with saturable absorber, there are parameters’ domains where synchronized classical laser solitons exist. They present localized lasing with frequency of the holding radiation. The synchronization domains are located outside the regions of bistability of homogeneous lasing modes. To synchronize the soliton, holding radiation with a very low intensity is sufficient. Note that this stage is important not only on its own but also for the statistical description of the system since we follow the quantum fluctuations against classical soliton.

The resulting equation enables us to study the laser soliton with or without the quantum features in it. To ensure the reliability of the results in quantum theory, we must very carefully define the Langevin sources. First, we must build all Langevin sources at a basic level. In this case, each canonical variable will have own source, and each of these sources in the adiabatic approximation somehow forms the resulting source in the equation for the generation field.

As for quantum features, they can arise for two reasons. Firstly, due to external reasons, namely, due to the sub-Poissonian nature of the excitation of an active (or even passive) medium. Second, due to internal reasons related to the complex nonlinear interaction of the field and matter. In the future, we intend to clarify these issues on the basis of the equation obtained in this article.

\acknowledgments

This work was supported by RFBR (grants 18-02-00402a and
19-32-90059) and within the Programme of the Russian Academy of Sciences "Modern methods of mathematical modelling in study of nonlinear dynamical systems".

\appendix

%Create the reference section using BibTeX:
\bibliography{LPh_QSoliton_2019}
%multiplr defined

%\begin{thebibliography}{100}
%
%\bibitem{WE}WE.
%
%\bibitem{SCULLY}SCULLY.
%
%\bibitem{DAVID} L. Davidovich. Rev. Mod. Phys. 68, 127 (1996).
%
%\end{thebibliography}

\end{document}